\documentclass[final,authoryear,5p]{elsarticle}

\usepackage{epsfig}

\usepackage{amssymb}

\usepackage{natbib}
\usepackage{amsmath}
\usepackage{numcompress}

\usepackage[ps2pdf,%
a4paper=true,%
breaklinks=true,%
colorlinks=true,%
pdfauthor={First Author et al.},%
pdftitle={Template for manuscripts in Advances in Space Research}%
]{hyperref}

\def\figfig#1#2#3{\centerline{\epsfig{figure=#2.eps,height=#1}}
\caption{\small #3}\label{fig:#2}}

\journal{Advances in Space Research}

\begin{document}

\begin{frontmatter}



\title{Accelerating NLTE radiative transfer by means of the Forth-and-Back
Implicit Lambda Iteration: A two-level atom line formation in 2D Cartesian coordinates}


\author{Ivan Mili\'{c}\corref{cor}}
\address{Astronomical observatory Belgrade, Volgina 7, 11060 Belgrade, Serbia \\
J. L. Lagrange Laboratory, UMR 7293, Universit\'{e} de Nice Sophia
Antipolis, CNRS, Observatoire de la C\^{o}te d'Azur, Campus
Valrose, 06108 Nice, France} \cortext[cor]{Corresponding author}
\ead{milic@aob.rs}


\author{Olga Atanackovi\'{c}}
\address{Department of Astronomy, Faculty of Mathematics, University of Belgrade, Studentski trg 16, 11000 Belgrade, Serbia}
\ead{olga@matf.bg.ac.rs}

\begin{abstract}

State-of-the-art methods in multidimensional NLTE radiative
transfer are based on the use of local approximate lambda operator
within either Jacobi or Gauss-Seidel iterative schemes. Here we
propose another approach to the solution of 2D NLTE RT problems,
Forth-and-Back Implicit Lambda Iteration (FBILI), developed
earlier for 1D geometry. In order to present the method and
examine its convergence properties we use the well-known instance
of the two-level atom line formation with complete frequency
redistribution. In the formal solution of the RT equation we
employ short characteristics with two-point algorithm. Using an
implicit representation of the source function in the computation
of specific intensities, we compute and store the coefficients of
the linear relations $J = a + bS$ between the mean intensity $J$
and the corresponding source function $S$. The use of iteration
factors in the 'local' coefficients of these implicit relations in
two 'inward' directions, along with the update of the source
function in other two, 'outward', directions leads to four times
faster solution than the Jacobi's one. Moreover, the update made
in all four consecutive sweeps of the grid leads to an
acceleration by a factor of 6-7 compared to the Jacobi iterative
scheme.
\end{abstract}

\begin{keyword}
radiative transfer; line formation; numerical techniques
\end{keyword}

\end{frontmatter}

\parindent=0.5 cm

\section{Introduction}

Radiative transfer (RT) is at the heart of many
astrophysical problems. In order to interpret the observed
spectra of astrophysical objects it is essential to solve the RT
problem. Radiation not only carries the information on the
physical state of the medium but also determines its structure and
properties. Above all, it plays a fundamental role in the
energy and force balance within the medium. Hence the need to
take it into account in modern 3D (magneto)hydrodynamic
simulations \citep[see, e.g.][]{Hayek10}. NLTE RT problems are
very demanding because of their non-local nature: radiation is
decoupled from the \emph{local} thermal state of the gas via
scattering processes, so that the state of the gas at one point in
the medium depends, via radiative processes, on the state of the
gas at all other points. In order to compute emergent intensity in
spectral lines (or, in general, the whole set of Stokes
coefficients) from a given atmospheric model (with a given run of
temperature and pressure/density), the coupled equations of
radiative transfer and statistical equilibrium have to be solved.
The coupling of the atomic level populations and the radiation
fields in the corresponding spectral line transitions is generally
highly non-linear. Because of all that, the specific
intensity of radiation, which fully describes the radiation field,
is a function of seven variables: three spatial and two angular
coordinates, frequency and time. Even if we neglect the time
dependence for the line transfer problems, and if we use a
discretization of all the variables with a grid of $100$ points
for each of them, we have specific intensity characterized by
$10^{12}$ values. Thus, the solution of the RT problem is very
time and memory consuming.

Due to these difficulties, NLTE RT problems have usually been
restricted to 1D geometry. However, for many objects (e.g.
inhomogeneous stellar atmospheres, rotating stars, accretion
disks, solar prominences) the plane-parallel or spherically
symmetric 1D approximation is inadequate. Although the theoretical
formulation of the multidimensional problem does not differ too
much from 1D case, the computational cost is increased by
many orders of magnitude. The direct solutions involving the
inversion of huge matrices are rather costly, while the most
simple iterative procedure, so-called $\Lambda $
iteration\footnote{$\Lambda$ operator was firstly introduced by
Schwarzschild as the operator acting on the source function to
give the mean intensity.}, that solves the problem equations in
turn, is usually too slow to be of practical use \citep[for
discussion on its convergence properties see,
e.g.][]{Mihalasbook}. Thus only fast iterative algorithms enable
efficient solution of multidimensional NLTE RT problems with the
short characteristics (SC) method almost exclusively used for the
formal solution. \citet{MAM78} were the first to apply SC
technique for the solution of RT in 2D slab geometries by using
difference approximation of the second-order differential
equations. \citet{KA88} developed an algorithm for the formal
solution based on SC solution of the first order differential RT
equations and parabolic approximation of the source function. This
SC technique was widely exploited in the last three decades within
so-called ALI (Accelerated Lambda Iteration) methods, based on the
operator perturbation technique \citep[for a review
see][]{Hubeny03}. Probably the most commonly used ALI method
is Jacobi iteration scheme that employs the diagonal (local) part
of the exact $\Lambda $ operator as an approximate lambda operator
(ALO) and computes the error caused by this approximation
iteratively \citep{OAB86}. It has been extended to NLTE line
transfer in 2D \citep[see e.g.][]{KO88, AP94,vanNoort02}, and to
polarized line RT: in 1D \citep{Faurobert97}, in 2D cylindrical
geometry \citep{Milic13}, and in 3D with partial frequency
redistribution (PRD) taken into account \citep{AN11}. The
convergence rate of the Jacobi method was usually increased by the
Ng acceleration technique \citep{Ng74}. The Gauss-Seidel method is
twice as fast as the Jacobi method, being usually further
accelerated by successive overrelaxation (SOR) technique
\citep{TBFB95}. It was generalized to the 2D line transfer problem
by \citet{LCP07}. Another very fast approach is bi-conjugate
gradient method \citep[e.g.][]{Papkalla95} that has been recently
generalized to multidimensional polarized line transfer with PRD
by \citet{ANP11}.

Here we propose another approach to the solution of 2D NLTE
radiative transfer problems. Our aim is to generalize to 2D
geometry the Forth-and-Back Implicit Lambda Iteration - FBILI,
previously developed for NLTE line transfer problems in 1D in the
paper by \citet{ACS97},
hereinafter ACS97. For simplicity, in this paper we
shall use two-level atom model. The multilevel atom case in 1D is
considered by {ACS97} and the transition from 1D to 2D will be
described in a forthcoming paper. FBILI is an extremely fast method, which without additional acceleration
technique significantly outperforms available methods in 1D problems \citep[for its convergent properties and
the problems solved, see][]{Atana07}. A very fast convergence to
the exact solution is achieved by the iterative computation of
the coefficients of implicit linear relations between the in-going
radiation field intensities and the line source function during
the forward sweep of the 1D grid and by their use in updating the
source function together with the specific intensities during the
backward sweep. Moreover, the use of an iteration factor in the
"local" coefficient of the implicit linear relations enormously
increases the convergence rate (for details see Section 2).

We recall the basic idea of FBILI method in the solution of NLTE
line formation in 1D geometry in Section 2. The implementation of
FBILI method to 2D Cartesian geometry is described in Section 3.
In Section 4 we solve a simple test problem and discuss the
results, and in Section 5 we comment on our future work.

\section{Forth-and-Back Implicit Lambda Iteration (FBILI) basics}

The FBILI method is developed and fully described in the paper by
ACS97. The essential features of this approach are the following:
\begin{itemize}
\item Two-point boundary nature of the problem, i.e. the existence of two separate
families of boundary conditions naturally suggests the separate
description of the propagation of the in-going intensities of the
radiation field $I^-_{\nu \mu }$ with initial conditions at the
surface and that of the out-going intensities $I^+_{\nu \mu }$
with initial conditions at the bottom of the system. This recalls the basic idea of a forth-and-back scheme.
\item The physics of radiative transfer is almost linear, hence a
linear algorithm is feasible for the solution of the problem.
\item An implicit representation of the source function is used in the
computation of both the in-going and the out-going intensities
with a piecewise parabolic behavior of the source function as a
suitable assumption.
\end{itemize}

Before we present the FBILI algorithm in more detail let us stress
here the main reason for its high convergence. Slow convergence of
the classical Lambda iteration is due to the fact that it computes
the \emph{total} mean intensity $J(\tau )$ from the old source
function $S^o(\tau )$, keeping thus from the previous iteration
more information than necessary. On the contrary, apart from 
the two-stream representation of the radiation field, in FBILI
$J(\tau )$ is split into a local and non-local component, with the local
part linearly dependent on the unknown local values of the source
function $S(\tau )$ and its derivative $S'(\tau )$. Only the
non-local part of the in-going mean intensity $J^-(\tau )$ is
computed from the old values $S^o(\tau )$ in the forward step,
whereas the non-local part of $J^+(\tau )$ and the local part of
both $J^-(\tau )$ and $J^+(\tau )$ are computed from the updated
values of $S(\tau )$ in the backward step. The fact that the only
peace of information transferred from the previous iteration is
contained in the non-local part of the in-going mean intensity
$J^-(\tau )$ enables an extremely high convergence rate of the
FBILI method.

In order to demonstrate FBILI approach, we shall consider the
two-level atom line transfer with complete frequency
redistribution in a static and isothermal plane-parallel 1D medium
with no background continuum. Under these assumptions, the RT
equation takes the form:
\begin{equation}
\mu {{dI_{\nu \mu }(\tau )}\over {d\tau }} = \phi _\nu [I_{\nu \mu
}(\tau ) - S(\tau )]\ ,\label{rte1}
\end{equation}
where $I_{\nu, \mu}(\tau )$ is the specific intensity of the radiation field at the
mean optical depth $\tau $, at frequency $\nu $ and direction $\mu
$ ($\mu $ is the cosine of the angle between the photon's
direction and the outward normal). The absorption-line profile,
$\phi _\nu $, is normalized to unity. The frequency independent line source function is
\begin{equation}
S(\tau )=\varepsilon B + (1-\varepsilon ) J(\tau ),\label{source}
\end{equation}
where $\varepsilon $ is the photon destruction probability, $B$ is
the Planck function, and 
\begin{equation} J(\tau ) = {1\over 2}\int _{-\infty }^{\infty} \phi _\nu d\nu \int _{-1}^1 I_{\nu \mu}(\tau )d\mu 
\label{scattering}
\end{equation}
is the scattering integral.

The specific intensities incident onto the boundaries, the
in-going intensities $I^-_{\nu \mu }(\tau =0)$ incident onto the
surface and the out-going intensities $I^+_{\nu \mu }(\tau =T)$
incident onto the bottom of the medium, are considered given.

In the numerical solution of the RT equation (\ref{rte1}) one
considers the discrete set of specific intensities with
frequencies $\nu _i$, $i=1,NF$ and directions $\mu _j$, $j=1,ND$,
and evaluates all the relevant depth-dependent functions on a
finite grid of mean optical depth values $\tau _l$, $l=1,NL$.

The propagation of the unknown radiation field "along a ray" can
be represented by using the integral form of the RT equation
\begin{equation}
I_{\nu \mu }(\tau _l) = I_{\nu \mu }(\tau _{l-1}) e^{-\Delta } +
\int _0^\Delta S(t)e^{t-\Delta } dt\ , \label{formal1}
\end{equation}
and adopting a polynomial representation for the source
function $S(\tau )$ between two successive depth points $l-1$ and
$l$. Here, $\Delta =\Delta \tau \phi _\nu /\mu $ is the
monochromatic optical path between the two points, with $\Delta
\tau = \tau _l - \tau _{l-1}$.

Assuming a piecewise parabolic behavior for the source function we can rewrite the RT equation (\ref{formal1}) for the in-going intensities in the following form:

\begin{equation}
I^-_l = I^-_{l-1} e^{-\Delta } + q^-_l S_{l-1} + p^-_l S_l + r^-_l
S'_l. \label{minus}
\end{equation}
Thus we get an implicit linear relation between the in-going
specific intensities and yet unknown local source function $S_l$
and its derivative $S'_l$. For brevity, in Eq. \ref{minus} we
omitted the dependence of $I$ on $\nu $ and $\mu $, and we put the
depth index as the subscript of all depth-dependent quantities.

The coefficients $p^-_l$, $q^-_l$ and $r^-_l$ depend only on the
optical distance $\Delta $. The first two terms on the right-hand
side of Eq.\,\ref{minus} represent the non-local part of the
in-going specific intensity, which is the only one that depends
linearly on the old values of the source function at all optical
depths $\tau <\tau _l$. The explicit values of $I^-_{l-1}$ are
obtained by previous recursive application of Eq. \ref{minus} with
the old values of $S(\tau )$ and $S'(\tau )$ at $\tau <\tau _l$.

Integrating Eq. \ref{minus} over frequencies and directions, we
get a local implicit linear relation:
\begin{equation}
J^-_l = a^-_l + b^-_l S_l + c^-_lS'_l.\label{Jminus}
\end{equation}
Proceeding from the given upper boundary condition for the
in-going intensities at the surface, $I^-_1$ (usually taken to be
zero), we compute the coefficients $a^-_l$, $b^-_l$ and $c^-_l$ at
all subsequent depth points $l>1$ to the bottom, and store them
for further use in the backward process of computation of the new
values of $S(\tau )$.

In the backward process, using the integral form of the RT
equation for the out-going intensities we can write

\begin{align}
I^+_l = & I^+_{l+1} e^{-\Delta } + \int _0^\Delta S(t)e^{t-\Delta } dt = \nonumber \\ 
& {I^+_{l+1} e^{-\Delta } + q^+_l S_{l+1}} +
p^+_l S_l + r^+_l S'_{l+1}.\label{plus}
\end{align}
Here again we assume piecewise parabolic behavior of the source
function within each layer $(\tau _l, \tau _{l+1})$.

We start from the bottom layer where the out-going specific
intensities $I^+_{NL}$ are given, and consequently $J^+_{NL}$ is
also known.

By taking into account Eq. \ref{Jminus} for $J^-_{NL}$, we derive
a similar relation for $J_{NL}$, from which, after we have
eliminated the derivative $S'_{NL}$ according to

\begin{equation}
S'_{NL-1}=S'_{NL}=[S_{NL}-S_{NL-1}]/\Delta \tau ,\label{linear}
\end{equation}
we obtain the coefficients $a_{NL}, b_{NL}$ and $c_{NL}$ of the
linear relationship

\begin{equation}
J_{NL} = a_{NL} + b_{NL}S_{NL} + c_{NL}S_{NL-1}.\label{Jnl}
\end{equation} 
On the other hand, Eq.\,\ref{Jminus} and angle- and line frequency integrated Eq.\,\ref{plus}, applied to the point $l=NL-1$, together with Eq.\,\ref{linear} allow us to express $J_{NL-1}$ also as a
linear combination of $S_{NL}$ and $S_{NL-1}$ with the known
coefficients:

\begin{equation}
J_{NL-1} = a_{NL-1} + b_{NL-1}S_{NL} +
c_{NL-1}S_{NL-1}.\label{Jnl1}
\end{equation} 
Substituting Eqs. \ref{Jnl} and \ref{Jnl1} into Eq. \ref{source} for $\tau _{NL}$ and $\tau
_{NL-1}$, respectively, we can easily derive the new values of
$S_{NL}$ and $S_{NL-1}$. The derivatives $S'_{NL}$ and $S'_{NL-1}$
are obtained from Eq. \ref{linear}, and the out-going
intensities $I^+_{NL-1}$ from Eq. \ref{plus}.

Let us note that when we solve RT problem in a constant
property, semi-infinite medium (as usual test problem), we take
that $J^+_{NL} = S_{NL}$ and $S'_{NL} = 0$, hence immediately
updating the source function according to:

\begin{equation}
S_{NL} = {{\varepsilon B + (1-\varepsilon ) a_{NL}^-}\over
{1-(1-\varepsilon )(b_{NL}^- +1)}}.
\end{equation}
For each successive upper depth point we proceed as follows.
The coefficients of the relation for $J^-_l$ (Eq. \ref{Jminus})
are known from the forward process. Since we assume parabolic
behavior of the source function, we can use the relation

\begin{equation}
S'_{l}={2\over {\Delta \tau }}[S_{l+1}-S_{l}] - S'_{l+1},
\label{elimination}
\end{equation}
to express $S'_l$ in terms of the known values of $S_{l+1}$ and
$S'_{l+1}$ and the thus far unknown $S_l$. Using Eq.\,\ref{elimination} we can
eliminate the derivative $S'_l$ from Eq.\,\ref{Jminus} to get $J^-_l$ as a linear function of $S_l$ only. Integrating the
formal solution for $I^+_l$ (Eq. \ref{plus}) and taking into
account that all the terms except $S_l$ are known, similar
expression for $J^+_l$ is straightforwardly derived. Consequently,
for each depth point $\tau _l$ we obtain the linear relation

\begin{equation}
J(\tau ) = a + b S(\tau )\label{ab1d}
\end{equation} 
that, together with Eq. \ref{source}, allows us to derive new value of $S_l$.
With new source function $S_l$ we can compute new derivative
$S'_l$ using Eq. \ref{elimination} and $I^+_l$ using Eq.
\ref{plus}. So, the computation of the new source function
together with the outgoing intensities is performed during the
backward process layer by layer to the surface.

Let us stress here that the iterative computation of the
coefficients of the implicit relations rather than that of the
intensities themselves, provides a high convergence rate. A much
higher convergence rate is achieved by the use of the iteration
factor $({I^-_{l-1}e^{-\Delta} + q^-_lS_{l-1}})/{S^o_l}$ in the
"local" coefficient (coefficient of the local source function $S_l$) of
Eq.\,\ref{minus}:

\begin{equation}
I^-_l = ({{I^-_{l-1}e^{-\Delta} + q^-_lS_{l-1}}\over
{S^o_l}}+p^-_l) S_l + r^-_lS'_l\ .\label{faktor}
\end{equation}
In other words, during the forward process at each depth
$\tau _l$ we retain, for further use in the back-substitution, the
ratio of the non-local part of the in-going intensity to the value
of the current local source function $S^o_l$. It represents
the only piece of information transferred from the previous
iteration. This ratio of two homologous quantities is a good
quasi-invariant iteration factor, which plays a very important
role in accelerating the iterative procedure. It quickly attains
its exact value and leads to the exact solution of the whole
procedure with an extremely high convergence rate.

\section{FBILI method in 2D}

In this Section we shall describe how FBILI method can be
implemented in the case of 2D medium in Cartesian geometry.

For simplicity we shall consider again two-level atom line
transfer with complete frequency redistribution in a static
isothermal medium with no background continuum. Let us assume that
the medium is infinite and homogeneous in the $z$-direction, so
that we solve the RT equation in the $(x,y)$ plane (see
Fig.~\ref{fig:ray_2n}) in the 'along the ray' form:

\begin{equation}
\frac{dI(x, y, \theta, \varphi, \nu)}{d\tau_s} = \phi(\nu )[I(x,
y, \theta, \varphi, \nu) - S(x, y)].
 \label{RTE}
\end{equation}
It is assumed that the object is represented by a 2D irregular
rectangular grid with $NX$ points in the $x$-direction and $NY$
points in the $y$-direction. The direction of propagation of the
photons is given by the polar angle $\theta $, measured with
respect to the $z$-axis, and the azimuthal angle $\varphi $,
measured with respect to the $x$-axis. The normalized line
absorption profile $\phi (\nu )$ for pure Doppler-broadening is
given by the Gaussian profile function $\phi (\nu ) =
{\frac{1}{{\sqrt {\pi }\Delta \nu _D}}} e^{-{(\nu -\nu
_0)}^2/{\Delta \nu _D}^2}$, and $d\tau _s$ is the line integrated
optical path length along the ray.

\begin{figure}[h]
\figfig{5.5cm}{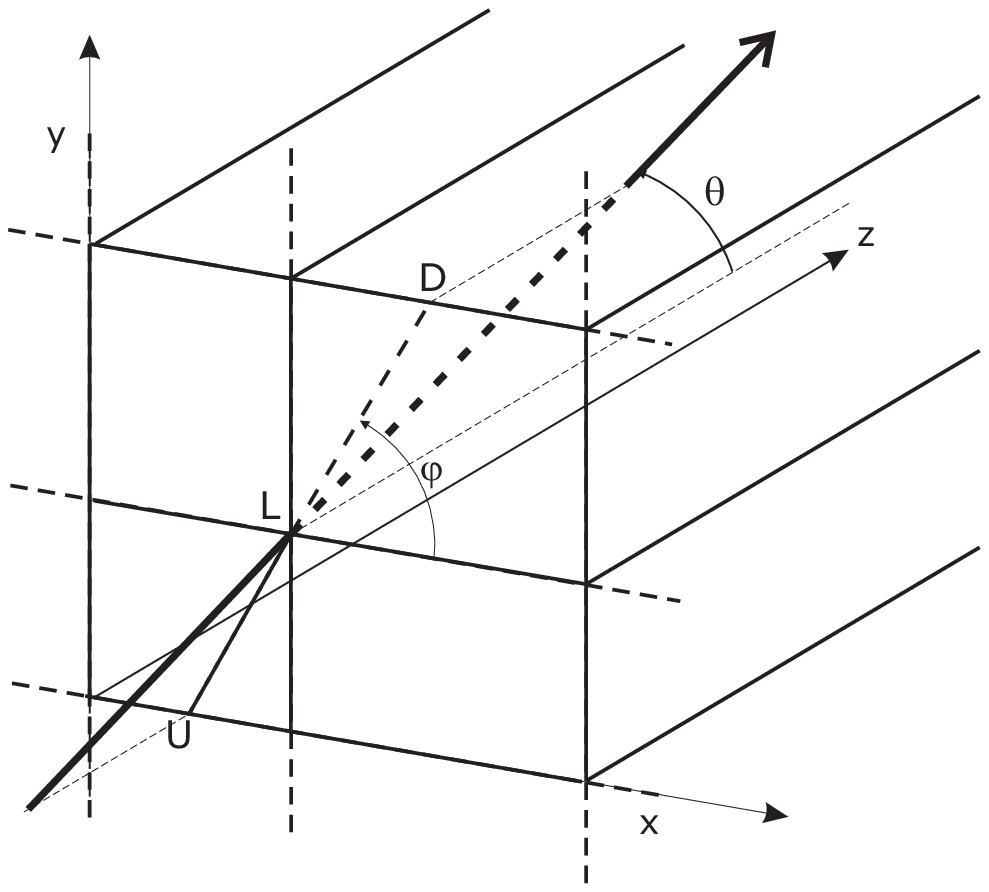}{Ray propagation and the definition of angles
in 2D geometry. The short characteristics at grid point $L$ for a
ray propagating from the lower left intersects the cell boundaries
at upwind point $U$ and downwind point $D$.}
\end{figure}

The two-level atom line source function in 2D is given by:

\begin{multline}
 S(x,y) = \varepsilon B + (1-\varepsilon)J(x,y)\\
 =\varepsilon B + (1-\varepsilon) \frac{1}{4\pi} \int_{-\infty}^{\infty}
 \phi(\nu) d\nu \oint I(x,y,\theta,\varphi,\nu)d\Omega ,
\label{SE}
\end{multline}
where $d\Omega = \sin{\theta}d\theta d\varphi$.

Here we shall describe how we can solve the problem equations
(\ref{RTE}) and (\ref{SE}) using the basic ideas of FBILI. Since
the formal solution of the RT equation is at the heart of each
iterative method we shall first present it as given by {ACS97}.

\subsection{Formal solution}

\begin{figure}
\includegraphics[width=0.49\textwidth]{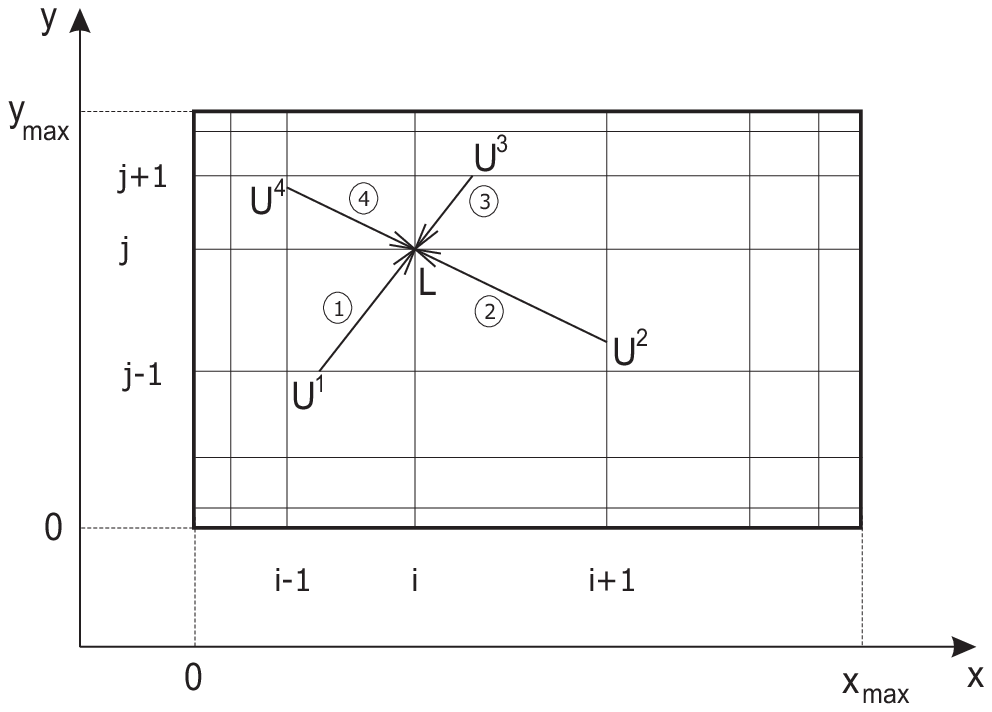}
\caption{Four sweeps through the local point
$L$ of 2D grid in $(x,y)$ plane. Short characteristics and the
corresponding upwind points $U^1$ - $U^4$ are indicated.}
\label{sweeps}
\end{figure}

In 2D geometry the formal solution of the RT equation is obtained
by sweeping the grid four times. We denote by $k(=1,2,3,4)$ the
directions of four sweeps in the corresponding quadrants of the
$x-y$ coordinate system (see Fig.\,\ref{sweeps}). Thus 1 denotes
the sweep in the direction of increasing $x$ and $y$, 2 - in the
direction of decreasing $x$ and increasing $y$, 3 - in the
direction of decreasing $x$ and $y$, and 4 - in the direction of
increasing $x$ and decreasing $y$. We take that $y=0$ is the
surface of the medium closer to the observer and we denote the
directions 1 and 2 as "inward" and the directions 3 and 4 as the
"outward" ones.

Like in most of the contemporary
methods, we use the integral form of the radiative transfer equation for
its formal solution and the so-called short characteristics
approach. In 2D geometry, for each sweep we can rewrite
Eq.\,\ref{formal1} in the following form:

\begin{equation}
I_L = I_U e^{-\Delta } + \int _0^{\Delta } S(t)e^{t-\Delta } dt.
\label{formal2d}
\end{equation}
For simplicity, here we drop index $k$ denoting the sweep
because all variables, except $S(\tau )$, are direction (sweep)
dependent. Here $\Delta $ is the monochromatic optical path
between the local grid point $L=(i,j)$ (i.e. point of interest, in
which the specific intensity is to be computed) and the "upwind"
point $U^k$, which is the nearest previous intersection point of
the direction of propagation of radiation $k$ and the grid lines
(see Figs.~\ref{fig:ray_2n} and \ref{sweeps}). The integral in
Eq.\,\ref{formal2d} can be solved analytically if we assume some
polynomial representation of the source function on each given
subinterval. In the standard short characteristics approach
\citep[e.g.][]{KO88}, assuming Lagrangean parabolic approximation,
the integral is expressed in terms of the source functions at
three points: upwind ($U$), local ($L$) and downwind ($D$) (the
latter being the successive intersection point, see
Fig.~\ref{fig:ray_2n}), so that Eq. \ref{formal2d} becomes:

\begin{equation}
I_L = I_U e^{-\Delta } + (\Psi _US_U + \Psi _LS_L + \Psi _DS_D),
\end{equation}
where the coefficients $\Psi $ follow from the interpolation
weights. Instead, for the formal solution FBILI method uses short
characteristics at two points, $U$ and $L$, expressing the
integral in terms of the source function at these two points and
the source function derivative at local point $L$:

\begin{equation}
I_L = I_U e^{-\Delta } + p_L S_L + q_L S_U + r_L S'_L.
\label{formal}
\end{equation}
It is important to note that upwind point $U$ \emph{is not} the
grid point and that the corresponding values of intensity and
source function, $I_U$ and $S_U$, must be evaluated by
interpolation \citep[see e.g.][]{AP94}. The
coefficients $p_L$, $q_L$ and $r_L$ depend solely on $\Delta $,
and thus implicitly on direction and frequency. If we assume
a piecewise parabolic behavior of the source function, their
values are easily computed from:

$$p_L=1-{2\over {\Delta
^2}}+e^{-\Delta }({2\over \Delta }+{2\over {\Delta ^2}})$$
$$q_L={2\over {\Delta
^2}}-e^{-\Delta }(1+{2\over \Delta }+{2\over {\Delta ^2}})$$
$$r_L=-1+{2\over \Delta
}-e^{-\Delta }(1+{2\over \Delta }).$$
Let us note that the specific intensity $I$ and the first
derivative of the source function $S'$ are the functions not only
of coordinates (like $S$), but also of direction and frequency.
The local derivative of the source function over the optical path
length can be expressed in terms of partial derivatives with
respect to $x$ and $y$-axes, and in the case of unit opacity
($\chi =1$) can be cast into the form:

\begin{align}
 S'_L(\nu, \theta, \varphi) & = \frac{1}{\phi(\nu)}[(\frac{\partial S}{\partial x})_L\cos{\varphi}\sin{\theta} \nonumber \\
 & + (\frac{\partial S}{\partial y})_L\sin{\varphi}\sin{\theta}].
 \label{partials}
\end{align}
The angles $\theta $ and $\varphi $ are shown in Fig.1.

Using Eq. \ref{partials}, Eq. \ref{formal} can be written for each
sweep as follows:

\begin{align}
I_L & =  I_U e^{-\Delta } + p_L S_L + q_L S_U  \nonumber \\
& + r_{L,x}(\frac{\partial S}{\partial x})_L + r_{L,y} (\frac{\partial
S}{\partial y})_L, \label{formalA}
\end{align}
where the coefficients $r_{L,x}$ and $r_{L,y}$ follow directly
from the above definition of the coefficient $r_L$ and
Eq.\,\ref{partials}. Once the values of the specific
intensity and the source function at the upwind point are obtained
by interpolation and after computing the coefficients $p_L$, $q_L$, $r_{L,x}$ and $r_{L,y}$, the only values that remain to be computed are
the local partial derivatives of the source function with respect
to $x$ and $y$.

\subsubsection{Computation of the derivatives}

The partial derivatives at the local point are obtained by
numerical differentiation. Here we use the Lagrangian
interpolation of the second order in three successive points
centered at the local one, that is:

\begin{equation}
  (\frac{\partial S}{\partial x})_{i,j} = w_{i-1,j,x} S_{i-1,j} + w_{i,j,x} S_{i,j} + w_{i+1,j,x} S_{i+1,j} ,
\label{derivativex}
\end{equation}
and

\begin{equation}
  (\frac{\partial S}{\partial y})_{i,j} = w_{i,j-1,y}S_{i,j-1} + w_{i,j,y}S_{i,j} + w_{i,j+1,y}S_{i,j+1}.
  \label{derivativey}
 \end{equation}
The explicit expressions for the weights in Eq.\,\ref{derivativex}
are:
\begin{eqnarray}
w_{i-1,j,x} = \frac{(x_i - x_{i+1})}{(x_{i-1} - x_i)(x_{i-1} - x_{i+1})} \nonumber \\
w_{i,j,x} = \frac{1}{x_i - x_{i+1}} + \frac{1}{x_i - x_{i-1}} \nonumber \\
w_{i+1,j,x} = \frac{(x_i - x_{i-1})}{(x_{i+1} - x_i)(x_{i+1} - x_{i-1})}
\end{eqnarray}
The weights used in Eq.\,\ref{derivativey} have the same
form, except they depend on the discrete values of $y$. At the
boundaries of the grid, linear approximation is used. Let us note
that the local source function $S_{i,j}$ contributes to the local
partial derivatives, so that its weight can be summed up with the
coefficient $p_L$ in Eq.\,\ref{formalA}. In some iterative
procedures described in the next section this led to better
stability and the convergence rate of the method.

The formal solution given above will be implemented in
various iterative schemes described in the next section.

\subsection{Iterative procedures}

Let us recall again that the simplest iterative scheme, $\Lambda$
iteration, computes the mean intensity ($J=\Lambda S$) and the
source function ($S=S(J)$) in turn. In order to compute the
mean intensity at any grid point it is necessary to perform four
sweeps of the grid, i.e. to compute the specific intensities
(using Eq.\,\ref{formalA}) at all previous grid points along each
sweep with the old (known from the previous iteration) values of
the source function. Once the mean intensities at all grid points
are obtained, one can compute new source function using
Eq.\,\ref{SE}. Iterations are repeated until the convergence is
achieved. As already mentioned, this is an extremely slow
procedure because it transfers from one part of the iterative step
to the other more information than necessary. In what follows
we shall explain how $\Lambda $ iteration in 2D has been
accelerated up to now and how it can be further accelerated by our
approach. More specifically, we shall describe our implementation
of Jacobi and Gauss-Seidel methods, and two variants of the
FBILI procedure applied to 2D line transfer problem.

\subsubsection{Jacobi-type iteration}

An efficient way to accelerate $\Lambda $ iteration is to simplify
the full description of the RT process, i.e.\,to use an
approximate lambda operator (ALO), $\Lambda ^*$, instead of the
full (exact) $\Lambda $ one, accounting for an error introduced by
this approximation iteratively. Using "operator splitting"
(well-known from numerical analysis) in RT computations, the
formal solution of the RT equation can be written in the form:

\begin{equation}
J = \Lambda S = (\Lambda - \Lambda ^*)S + {\Lambda ^*}S.
\label{split}
\end{equation}
\citet{OAB86} were the first to point out that the diagonal of the
exact $\Lambda $ matrix itself represents an almost optimum ALO.

Here, we shall describe the Jacobi-type iterative procedure and
see that the coefficient of the local source function $b_L$ plays
a role of the diagonal ALO in the Jacobi method.

In the Jacobi-type procedure applied to 2D radiative transfer, first we have
to sweep the grid 4 times, and in every sweep $k$ to compute and
store the coefficients of the linear relation:

\begin{equation}
J^k_L=a^k_L+b^k_LS_L. \label{abi}
\end{equation}
This equation is obtained by the angle- and line profile
integration of Eq.\,\ref{formalA}, in such a way that the
coefficient $a^k_L$ contains all non-local contributions to the
specific intensity at the given point $L$:

\begin{align}
 a^k_L = & \frac{1}{4\pi}\int_{-\infty}^{\infty} \phi(\nu) d \nu \int \Bigl[I^k_U e^{-\Delta ^k} + q^k_L S^k_U + \nonumber \\
 &r^k_{L,x}(\frac{\partial S}{\partial x})^k_L + r^k_{L,y}(\frac{\partial S}{\partial y})^k_L\Bigr] d \Omega,
\end{align}
and is computed using the current values of the source function
and its derivatives, whereas the coefficient $b^k_L$ has the form:

\begin{equation}
b^k_L = \frac{1}{4\pi}\int_{-\infty}^{\infty} \phi(\nu) d \nu \int
p^k_L\, d \Omega, \label{b_def}
\end{equation}
playing the role of the diagonal ALO. To be consistent,
the contribution of the local source function to the local partial
derivatives (see Eqs\,\ref{derivativex} and \ref{derivativey})should be included in the coefficient $b_L^k$ rather
than in the coefficient $a_L^k$.

The total mean intensity at point $L$ is obtained by summing up
mean intensities in all the sweeps, and is given by

\begin{equation}
J_L=a_L+b_LS_L, \label{ab}
\end{equation}
where $a_L=\Sigma _{k=1}^4 {a^k_L}$ and $b_L=\Sigma _{k=1}^4
{b^k_L}$ are the total coefficients.

Once we know the coefficients of Eq.\,\ref{ab}, by inserting
Eq.\,\ref{ab} into Eq.\,\ref{SE} we can update the source function
at all depth points throughout the 2D grid by means of:

\begin{equation}
 S_L = \frac{\varepsilon B + (1-\varepsilon )a_L}{1 - (1-\varepsilon ) b_L}.
 \label{FBILI}
\end{equation}
In this way, the iterative computation of the coefficients $a_L$
and $b_L$ of the implicit relation (\ref{ab}) instead of the
unknown quantities ($J_L$ and $S_L$) themselves leads to much more efficient
corrections than in $\Lambda $ iteration. This scheme reduces
number of iterations by a few orders of magnitude with respect to
the ordinary $\Lambda$ iteration. However, even this is not fast
enough for some more demanding problems (strong, scattering
dominated lines).

\subsubsection{Gauss-Seidel-type iteration}

As it has just been explained, in the Jacobi iteration the
grid is swept four times and in every sweep the coefficients
$a^k_L$ and $b^k_L$ are computed from the "old" values of the
source function. Only after getting the total coefficients $a_L$ and $b_L$ \emph{at all grid
points}, the source function is updated using Eq.\,\ref{FBILI}.

The Jacobi scheme can be substantially accelerated if the new
source function is computed \emph{as soon as the total
coefficients $a_L$ and $b_L$ in Eq.\,\ref{ab} are available
(known)} at some point. This is, for example, the situation at the
boundary grid points after sweeping the grid three times and
computing the corresponding coefficients $a^k_L$ and $b^k_L$
($k=1,3$). We start the fourth sweep with given values
of $a^4_L$ and $b^4_L$ at two boundaries: $(1,j);\, {j=1,NY}$ and
$(i,NY);\, {i=1,NX}$ (see Fig.\,3). The new source function $S_L$ at these points is easily computed
using Eq.\,\ref{FBILI}. Now, our aim is to come up with the scheme which will use this idea at all subsequent points as
the use of "new" (updated during the current sweep)
source functions in the computation of the local intensities in
the fourth sweep accelerates the convergence. This numerical
scheme corresponds to Gauss-Seidel method known from numerical
algebra \citep[see e.g.][]{Saad}. For the solution of the 1D NLTE
RT problem this idea was implemented in two different ways by
\citet{TBFB95} and \citet{ACS97}. In the paper by \citet{TBFB95}
standard approximate $\Lambda$ operator approach with three-point
algorithm to set up short characteristics of the second order is
used. This method has been explicitly generalized to 2D geometry
by \citet{LCP07}. The FBILI method, developed by \citet{ACS97},
uses two-point algorithm and computes the coefficients of the
implicit relations expressing the intensities in terms of the
source functions and its derivatives at pairs of successive depth
points.

The whole procedure is more complicated in multidimensional
geometries because of the spatial interpolations needed to obtain
values of the upwind source function and intensities. Let us
consider the procedure in 2D in more detail.

\begin{figure}
\includegraphics[width = 0.49\textwidth]{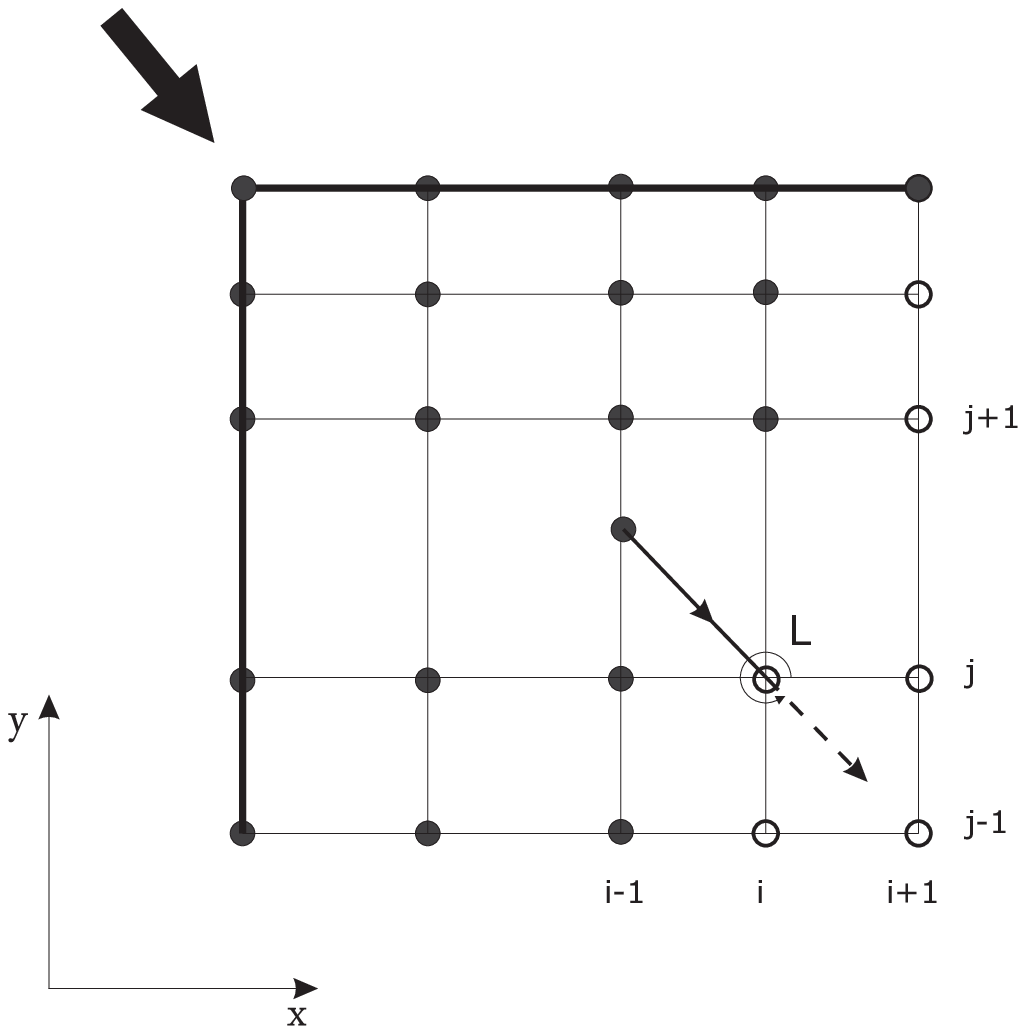}
\caption{2D grid sweep in the 4th direction.
Full dots correspond to the new values of the source function,
empty ones to the old values.}
\label{4_sweep}
\end{figure}

Fig.\,\ref{4_sweep} describes the situation upon arrival at the grid
point $(i, j)$ in the last, fourth sweep, after the 2D grid was
swept three times. We assume that the source function is already updated in the points represented by full dots. From now on we
shall refer to the sweeps during which the formal solution is
performed, and appropriate coefficients are stored , with no update of the source function as the
\emph{forward} sweeps, whereas the sweeps during which the source
function is updated as the \emph{backward} ones \footnote{For example,
Jacobi iteration consists of four forward sweeps followed by the
simultaneous update of the source function over the entire grid.}.

It is essential to realize that all the non-local
contributions to the coefficient $a_L$ in Eq.\,\ref{FBILI} must be
properly taken into account. Some of these contributions are
already updated in the current sweep ("new"), while the others
still have their values from the previous iteration ("old").

In our implementation of Gauss-Seidel iterative scheme we first modify the formal solution in the following way: We use Eq.\,\ref{formalA} with partial
derivatives given by Eqs.\,\ref{derivativex} and
\ref{derivativey}, thus expressing explicitly the contributions of
$S_{i-1,j}$, $S_{i+1,j}$, $S_{i,j-1}$ and $S_{i,j+1}$ to the local
specific intensity. Furthermore, upwind source function $S_U$ is
also expressed in terms of the source function values at the
neighboring grid points. As an example, for the point U$_1$ in
Fig.\,\ref{sweeps} we have:

\begin{align}
S_U & = W_{i-1,j-1} S_{i-1,j-1} + W_{i,j-1} S_{i,j-1} \nonumber \\
& + W_{i+1,j-1} S_{i+1,j-1}.
\end{align}
In the above equation the weights $W$ follow from the
Lagrangean interpolation of the second order, and, for
convenience, we give the expressions:

\begin{eqnarray}
W_{i-1,j-1} = \frac{(x_U - x_i) (x_U - x_{i+1})} {(x_{i-1} - x_i) (x_{i-1} - x_{i+1})}, \nonumber \\
W_{i,j-1} = \frac{(x_U - x_{i-1}) (x_U - x_{i+1})} {(x_{i} - x_{i-1}) (x_{i} - x_{i+1})}, \nonumber \\
W_{i+1,j-1} = \frac{(x_U - x_{i-1}) (x_U - x_{i})} {(x_{i+1} - x_{i-1}) (x_{i+1} - x_{i})}.
\end{eqnarray}
Finally, Eq.\,(\ref{formalA}) for each sweep $k$ takes the new form:

\begin{equation}
 I^k_L = I^k_U e^{-\Delta ^k} + p^k_L S_L + \sum_{i'} \sum_{j'} r^k_{i',j'}S_{i',j'}.
 \label{explicit}
\end{equation}
The expression for the parabolic interpolation formula in the
above equation is similar to the one given by Eq.\,5 in the paper by
\citet{KA88}. Here the coefficients $r_{i',j'}$ follow from the
approximations used to compute the local derivative of the source function
$S'_L$ and
to interpolate the value of the source function at upwind point
$S_U$. Note that $r_{i,j}$ (indices $(i,j)$ refer to the grid point $L$)
is always zero, as all local contributions are added to the coefficient
$p_L$. This way, all
non-local contributions (in all sweeps) except the upwind specific
intensities are explicitly expressed using eight neighboring
source functions.

Integration of Eq.\,\ref{explicit} over angles and line profile yields:

\begin{equation}
 J_L^k = a_L^k + b_L^k S_L + \sum_{i'} \sum_{j'} c_{i',j'}^k S_{i',j'}\,.
 \label{explicit_integrated}
\end{equation}
Here, the coefficients are defined as:

\begin{equation}
a^k_L = \int \phi (\nu )d\nu \int \frac{d\Omega }{4\pi } I^k_U
e^{-\Delta ^k},
\end{equation}
\begin{equation}
b^k_L = \int \phi (\nu )d\nu \int \frac{d\Omega }{4\pi }\,p^k_L,
\end{equation}
and

\begin{equation}
c^k_{i',j'} = \int \phi (\nu )d\nu \int \frac{d\Omega }{4\pi}\,r^k_{i',j'}.
\end{equation}
After computing the coefficients $a^k_L$, $b^k_L$ and
$c^k_{i',j'}$ in all four directions, the source function can be
updated according to:

\begin{equation}
 S_L = \frac{\varepsilon B + (1-\varepsilon )(a_L + \sum_{i'} \sum_{j'} c_{i'j'} S_{i'j'} )}{1 - (1-\varepsilon )
 b_L},
 \label{FBILI_B}
\end{equation}
where $a_L = \sum_k a_L^k$, $b_L = \sum_k b_L^k$ and $c_{i'j'} = \sum_k c_{i'j'}^k$.

Let us point out here that the upwind intensity $I_U^4$, contained in
the coefficient $a_L^4$, is computed from the updated source function at
previous points along the fourth sweep. It is important to stress that if Eq.\,\ref{FBILI_B} is used to
update the source function in the backward sweep, all the proper
contributions of ``new'' and ``old'' neighboring source functions
are \emph{automatically} taken into account, through the sum
$\sum_{i'} \sum_{j'} c_{i'j'} S_{i'j'}$. We now propose the
following, Gauss-Seidel like scheme:

\begin{enumerate}
\item Sweep the grid in the first three directions (forward sweeps), computing and storing the corresponding
 coefficients $a_L^k$, $b_L^k$ and $c_{i',j'}^k$, $(k=1,3)$ of Eq.\,\ref{explicit_integrated} by means of the old values of the source function.
 \item Start the fourth (backward) sweep. At the grid points on the two boundaries (marked in bold in Fig.\,\ref{4_sweep}),
 specific intensities of
 the incident radiation field are known so that $a_L^4$, $b_L^4$ and $c_{i',j'}^4$ are known,
 and the source function $S_L$ can be straightforwardly computed using Eq.\,\ref{FBILI_B}. After updating the source function, specific intensity $I_L^4$ is computed using Eq.\,\ref{explicit}.
 \item At all the subsequent points of the backward sweep, with the updated values of the specific intensities $I^4_L$ at previous points, the upwind intensity $I^4_U$ is to be computed, and, hence the coefficient $a^4_L$. Once the total coefficients $a_L$, $b_L$ and $c_{i',j',L}$ are obtained, the source function is updated by means of Eq.\,\ref{FBILI_B} and specific intensity is computed using Eq.\,\ref{explicit}.
 \item Steps 1-3 are repeated until convergence.
\end{enumerate}

The main difference between this scheme and the above
described Jacobi-like scheme is that the source function is
updated \emph{in the course} of the fourth sweep (instead \emph{after} the
fourth sweep is completed). This modification introduced by the
Gauss-Seidel approach significantly increases the rate of
convergence. As we shall see in
the next section, even further acceleration in 2D is possible by the
application of the forth-and-back approach and the use of iteration
factors.

\subsubsection{``Two-by-two'' FBILI method}

FBILI method proposed by ACS97 brought about improvements over the
existing ones in the following: (i) iterative computation of the
coefficients of the implicit linear relation between the specific
intensities and the local source function and its derivative in
the forward sweep, combined with an efficient method of
back-substitution (a two-point, not a three-point scheme), led to
a quick update of $S$ and $S'$ along the 1D grid, and (ii)
the use of iteration factor in the forward sweep, which
``enhances'' the local operator by $(I_U e^{-\Delta} + qS_U) /
S_L$, provided an extremely fast convergence with respect to the
previous schemes. The acceleration of the iterative procedure is
due to the fact that it is much faster to iterate on the ratio of
the two unknowns than on the unknowns themselves. In 1D case,
introduction of the iteration factor increased the convergence
rate of the FBILI method by a factor of 3.

In order to generalize FBILI to 2D we ought to take into account
that, due to the twofold two-point boundary nature of the problem,
we have two pairs of the mutually opposite sweeping directions
(1-3 and 2-4). We can, therefore, emulate the original FBILI
approach by considering two inward directions (1 and 2) as the
forward ones and two outward directions (3 and 4) as the backward
ones. The update of the source function is thus performed twice
during the single iteration. Moreover, in the forward sweeps we
can introduce appropriate iteration factors into the 'local'
coefficient $b_L$ to speed up the convergence.

In general, the method can be used in many different ways: it is
possible to use iteration factors in one or two directions, or not
at all; there can be one, two, or even four backward sweeps. In
the following we present some of the most efficient schemes.

As before, we use Eqs. \ref{explicit} and
\ref{explicit_integrated}, and we include the iteration factors
in the computation of the coefficient $b_L$ during the two
in-going (forward) sweeps 1 and 2:

\begin{equation}
 b^{1,2}_L = \int \phi(\nu) d \nu \int (p^{1,2}_L + \frac{I_U^{1,2} e^{-\Delta ^{1,2}}}{S_L^{\rm{old}}}) d\Omega,
 \label{if}
\end{equation}
while in the out-going (backward) directions the coefficient
$b_L$ contains only direction- and line profile-integrated
coefficient $p_L$ from Eq. \ref{explicit}.

We propose the following iteration procedure:
\begin{enumerate}
 \item Sweep the grid three times (forward sweeps), computing the specific intensity using Eq.\,\ref{explicit} and iteration factors in directions 1 and 2 as given by Eq.\,\ref{if}.
 Compute the corresponding coefficients $a^{1-3}_L$, $b^{1-3}_L$ and
 $c^{1-3}_{i',j'}$.
 \item In the fourth (backward) sweep, starting from the grid points on two corresponding boundaries with known boundary conditions,
 update the source function by means of Eq. \ref{FBILI_B} and the out-going specific intensity using Eq. \ref{explicit} point by point throughout the grid.
 Reset coefficient $b^1$ to zero (as the iteration factor is used, $a^1$ is zero by default).
 \item Sweep the grid in direction 1 (iteration factor is used). Reset $a^3$ and $b^3$ to zero.
 \item Sweep the grid in direction 3 (no iteration factor). \emph{Update the source function and the intensity while performing the sweep}.
 Note that this sweep is now backward sweep. Reset $b^2$ to zero ($a^2$ is zero by default).
 \item Sweep the grid in direction 2 (iteration factor is used). Reset $a^4$ and $b^4$ to zero.
 \item Sweep the grid in direction 4 (no iteration factor). \emph{Update the source function and the intensity while performing the sweep.}
 Reset $b^1$ to zero.
 \item Repeat steps 3-6 until convergence.
\end{enumerate}

The only differences in the above scheme, with respect to our
GS-like procedure described in section 3.3.2 are: (a) inclusion of
factors in "in-going" directions 1 and 2, (b) a re-ordering of
directions (for better stability), and (c) updating of the source
function in "out-going" directions 3 and 4, i.e. there are two
backward sweeps now instead of just one. Hence, the source
function is updated twice per iteration, i.e. once per each pair
of the mutually opposite sweeping directions (1-3 and 2-4). This
implementation shows a very good stability and also much better
convergence properties with respect to other methods described
previously. This inspired us to try to further accelerate the
method by updating the source function \emph{in all four sweeps}.

\subsubsection{"Sweep by sweep" FBILI procedure}

In the previous section we have seen that the update of the source
function can be performed more than once during a single
iteration. This idea was realized in 1D
plane-parallel geometry by means of SSOR (symmetric successive
overrelaxation) method \citep[see, for example: ][]{SSOR10}. In
principle, as soon as the grid is swept four times in the first
iteration and coefficients $a^{1-4}$, $b^{1-4}$ and
$c^{1-4}_{i'j'}$ are known, one can update the source function in
\emph{every} sweep of the grid.

Here, the only difference with respect to the "two-by-two"
procedure is that after the step 2, source function is updated
during all four sweeps ("sweep by sweep"). This leads to four
updates per iteration at essentially no additional computational
cost (computation of the source function takes negligible time
with respect to the formal solution). This very same procedure
without iteration factors would correspond to Symmetric
Gauss-Seidel (SGS) in 2D geometry. As we shall see in the next
section, this method, with the help of iteration factors,
extremely accelerates the convergence with no additional numerical
acceleration technique. 

\section{Results}
In order to test the properties of the above mentioned procedures
we solve the problem given by \citet{AP94}. We consider a slab
with optical depth $\tau =10^4$ along both ($x$ and $y$) axes,
with $\varepsilon =10^{-4}$, $B=1$, and Doppler profile.
Equidistant logarithmic spacing in optical depth with
approximately 10 points per decade (129$\times$129 points) is
used. The slab is irradiated at bottom and at side boundaries,
from the angles $\pi<\varphi<2\pi$, with radiation equal to $B$.
For angular integration we use Carlson's set B \citep{Carlson63}
with $n=8$ (12 angles per octant). We use 9 frequency points in a
half of the line profile, and the trapezoid integration weights.

The properties of the iterative procedures are analyzed by
calculating at each iteration step $i$ the maximum relative change
of the solution between two successive iterations $i-1$ and $i$:

\begin{equation}
R_c^{i} = \vert {{S^{i} - S^{i-1}}\over {S^{i}}} \vert _{\rm max}\\ .
\end{equation}
The first tested procedure, denoted here as Jacobi-type procedure,
needed 118 and 195 iterations to reach the maximum relative change
$R_c=10^{-3}$ and $R_c=10^{-5}$, respectively. From these results
it is evident that higher convergence rate is desired. As
mentioned before, it is customary to apply Ng acceleration
\citep{Ng74} to Jacobi method. However, since it requires some
experimentation (its use is not straightforward), we have not used
it, i.e.\,we present here the results with no additional
mathematical acceleration techniques like the Ng's.

When we applied the second, Gauss-Seidel type procedure we
obtained the corresponding solutions in 77 and 126 iterations. The
increase in the convergence rate is evident, but not as great as
in 1D case.

\begin{figure}[h]
\figfig{5.5cm}{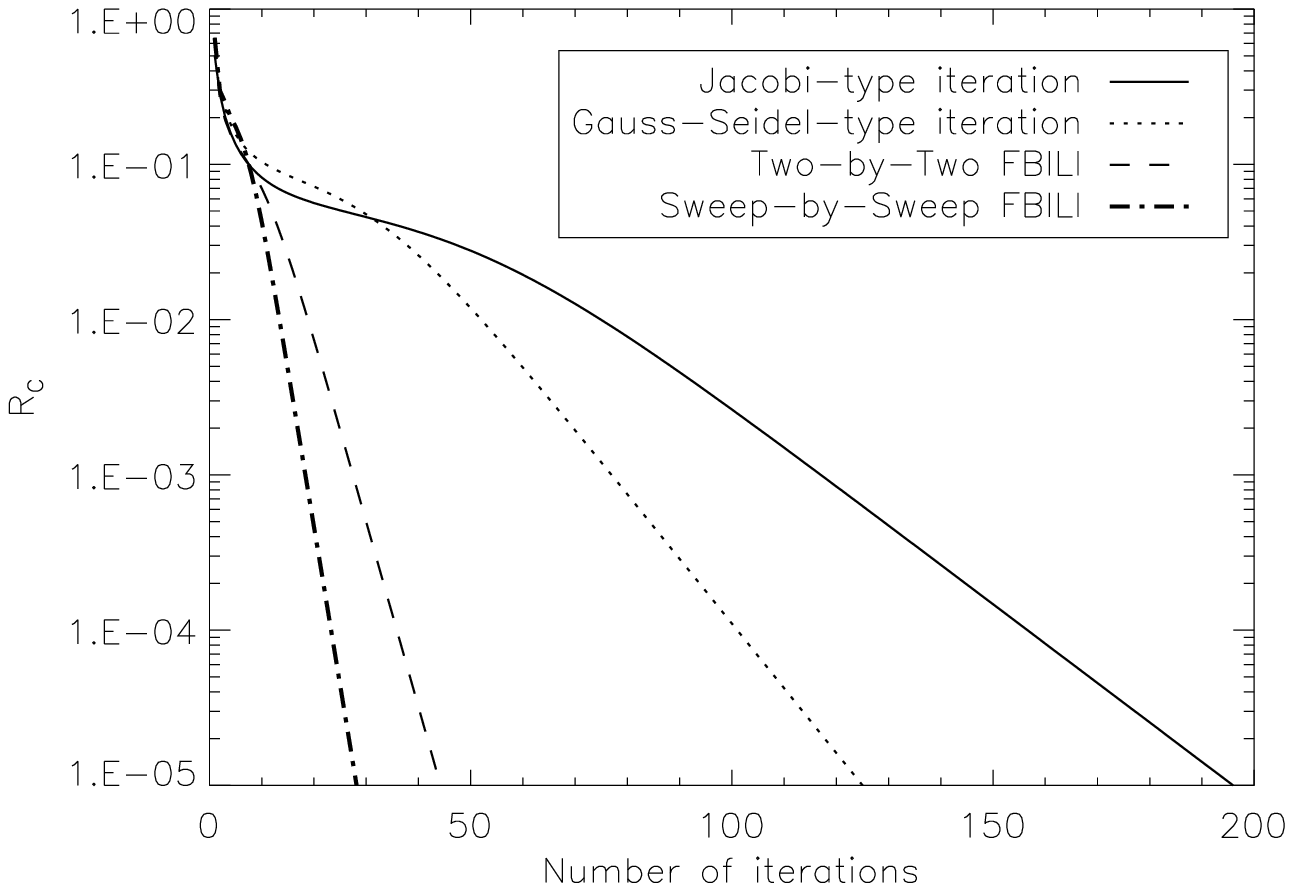}{Variation of the maximum relative change
with iterations for the iterative procedures considered.}
\end{figure}

Finally, two FBILI procedures with iteration factors ("2-by-2" and
"sweep-by-sweep") dramatically increased the convergence rate (see
Fig.~\ref{fig:relative_2}). In these two procedures, we define one
iteration as the whole set of four sweeps, although the source
function is updated two(four) times (recall that the computation
of the source function is very fast). "Two-by-two" FBILI with
iteration factors in directions 1 and 2 reaches $R_c = 10^{-3}$ in
28 iterations and $R_c = 10^{-5}$ in 44 iterations, while the
"sweep-by-sweep" procedure with iteration factors in two inward directions and
source function update in all four sweeps achieves the above
relative changes in 19 and 29 iterations, respectively (6-7
times faster than Jacobi scheme). Omitting the iteration factors
in the procedure that updates the solution in all four directions
leads to the iterative scheme corresponding to the generalization
of the Symmetric-Gauss-Seidel method. Using this procedure the
above convergence criteria are satisfied in 39 and 61 iterations,
respectively. The importance of the iteration factors is evident,
as they improve the convergence rate of SGS by a factor of
more than two.

In order to study the performance of an iterative method, also the
\emph{true} error ought to be analyzed. Since the analytical
solution of this benchmark problem cannot be obtained, the true
error is expressed with respect to $S^{\infty }_{\rm REF}$ - the
fully converged "exact" solution obtained with some well-tested
code. In this case we used the 1000th Jacobi iteration, with four
times more dense spatial grid (for which $R_c \approx 10^{-15}$)
as the "exact" solution. Considering that the source function
along the central line of the slab, $S(NX/2,j); j=1,NY$, has a
similar behavior to the solution in a 1D semi-infinite stellar
atmosphere, we took central surface point as the point of interest
in analyzing the true error. So, we define maximum relative true
error as:

\begin{equation}
T_e^{i} = \vert {{S(NX/2, 1)^{i} - S(NX/2,1)^{\infty }_{\rm REF}} \over
{S(NX/2,1)^{\infty }_{\rm REF}}}\vert _{\rm max} \ .
\end{equation}

\begin{figure}[h]
\figfig{5.5cm}{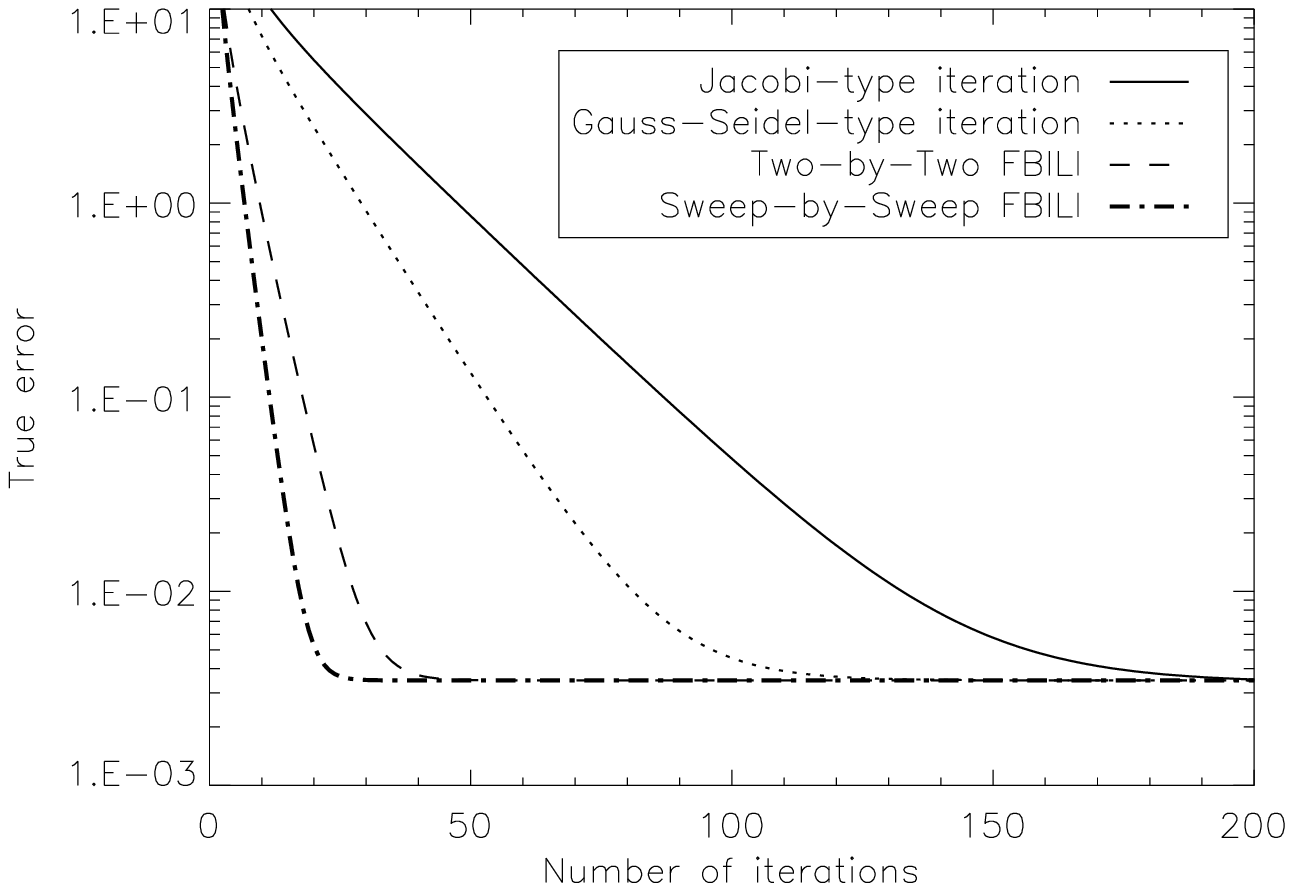}{Variation of the true error with iterations
for the procedures considered.}
\end{figure}

Change of the true error with number of iterations is shown in
Fig.\,\ref{fig:true_2}. Excellent properties of FBILI method are
again evident. For a fast converging method such as this one, one
can actually use weaker convergence criterion in terms of $R_c$.
To make this statement clear, recall that slowly converging method
will reach small relative change relatively quickly, but might
still be far from the ``true'' solution. We stress that, in
principle, the true error should be the convergence criterion, but
as it is not known, good knowledge of the convergence properties
of the method in question must be obtained in order to set proper
value of $R_c$ as the convergence criterion, thus optimizing the
computing time.

\section{Conclusions}

We have presented main concepts of a new iterative scheme for the
NLTE line radiative transfer in 2D Cartesian geometry.
Introduction of iteration factors in the 'local' coefficient of
the linear relation between $J$ and $S$, combined with the idea of
using new values of the source function as soon as they are
available in all four sweeps of the grid dramatically improves the
convergence rate.

Even better comparative convergence properties of FBILI method can
be expected in applications to some more realistic problems; e.g.
in the semi-infinite atmosphere with periodic boundary conditions.
Just as 1D FBILI iterative scheme shows its full advantage for
optically thick media and scattering dominated problems that need
fast methods to be solved efficiently, we aim to achieve the same
in the 2D and 3D cases. Our first goal is to implement periodic
boundary conditions in the code and also to test more accurate
interpolating strategies \citep[for example, cubic interpolation, as suggested by\,][]{Sim12} in computing formal solution, the source
function derivatives, and the spatial interpolation at the upwind
point. We also aim to generalize the backward elimination scheme from 1D FBILI to 2D, i.e.\,to eliminate the local derivative by means of the source function and the derivative at previous grid points. This would eliminate need for keeping eight $c_{i'j'}$ coefficients and lead to more elegant solution.

In the future work we will also test scaling of
convergence properties with respect to grid resolution and
demonstrate generalization of the method to multilevel atom case,
as well as to PRD problems and polarized line transfer in a
two-level atom approximation.

\section{Acknowledgements}

We thank Marianne Faurobert for useful discussions during IM's and
OA's stay in Nice. We are indebted to anonymous referees, not only
for useful comments on an earlier version of this manuscript, but
also for pointing us possible improvements of our method. This
research is being done in the framework of the project 176004,
"Stellar Physics", supported by the Serbian Ministry of Science
and Education.





\end{document}